\documentclass[12pt]{article}
\renewcommand{\thefootnote}{\fnsymbol{footnote}}
\newcommand{\beq}{\begin{equation}}
\newcommand{\eeq}{\end{equation}}
\newcommand{\beqn}{\begin{eqnarray}}
\newcommand{\eeqn}{\end{eqnarray}}
\newcommand{\beqlab}[1]{\begin{equation}\label{#1}}
\newcommand{\beqnlab}[1]{\begin{eqnarray}\label{#1}}
\newcommand{\mycite}[1]{\cite{#1}}
\newcommand{\eq}[1]{Eq.~(\ref{#1})}
\newcommand{\grad}{\mbox{\boldmath ${\nabla}$\unboldmath}}

\newcommand{\veps}{\mbox{\boldmath$\epsilon $}}

\newcommand{\vpi}{\mbox{\boldmath$\pi$}}
\newcommand{\vsig}{\mbox{\boldmath$\sigma$}}

\newcommand{\vk}{{\bf k}}
\newcommand{\vr}{{\bf r}}
\newcommand{\vR}{{\bf R}}
\newcommand{\vn}{{\bf n}}
\newcommand{\vS}{{\bf S}}
\newcommand{\vp}{{\bf p}}

\newcommand{\vx}{{\bf x}}
\newcommand{\vl}{{\bf l}}

\newcommand{\vq}{{\bf q}}

\newcommand{\vD}{{\bf D}}
\newcommand{\vM}{{\bf M}}
\newcommand{\vJ}{{\bf J}}
\newcommand{\vH}{{\bf H}}

\newcommand{\vs}{{\bf s}}

\newcommand{\vA}{{\bf A}}

\newcommand{\vP}{{\bf P}}
\newcommand{\bra}{\langle\psi_0|}
\newcommand{\ket}{|\psi_0\rangle}

\begin{document}

\begin{titlepage}

\vspace{1cm}
\begin{flushright}
{\bf Budker INP 99-76\\
January 22, 2001 }
\end{flushright}

\vspace{1cm}

\begin{center}
{\LARGE \bf Relativistic corrections to the electromagnetic
polarizabilities of compound systems}\\
\vspace{1cm}
R.N.~Lee$^a$\footnote{e-mail:R.N.Lee@inp.nsk.su},
A.I.~Milstein$^a$\footnote{e-mail:A.I.Milstein@inp.nsk.su}
M.~Schumacher$^{b}$\footnote{e-mail:Martin.Schumacher@phys.uni-goettingen.de}\\
{\it
$^a$Budker Institute of Nuclear Physics, 630090 Novosibirsk, Russia\\
$^b$Zweites Physikalisches Institut, Universit\"at G\"ottingen, D-37073
     G\"ottingen, Germany}\\
\end{center} \vspace{1cm} \begin{abstract} The low-energy
amplitude of Compton scattering on the bound state of two
charged particles of arbitrary masses, charges and spins is
calculated. A case in which the bound state exists due to
electromagnetic interaction (QED) is considered. The term,
proportional to $\omega^2$, is obtained taking into account the
first relativistic correction. It is shown that the complete
result for this correction differs essentially from the commonly
used term $\Delta\alpha$, proportional to the r.m.s. charge
radius of the system. We propose that the same situation can
take place in the more complicated case of  hadrons.
\end{abstract}
\end{titlepage}

\section{Introduction}
\renewcommand{\thefootnote}{\arabic{footnote}}
\setcounter{footnote}{0} The electromagnetic polarizabilities
$\bar\alpha$ and $\bar\beta$ are fundamental characteristics of
the bound system. Their magnitudes depend not only on the
quantum numbers of the constituents, but also on the properties
of the interaction between these constituents. Therefore, the
experimental and theoretical investigation of the
electromagnetic polarizabilities are of a great importance. In
particular, their prediction and the comparison with
experimental data may serve as a sensitive tool for tests of
hadron models. Correspondingly, a large number of researchers
have been attracted by this fascinating possibility. The
electromagnetic polarizabilities can be obtained from the
low-energy Compton scattering amplitude. In the lab frame the
amplitude of Compton scattering on the compound system of total
angular momentum $S=0,\,1/2$ up to $O(\omega^2)$ terms reads
\mycite{klein55,petrunkin61} \beqlab{T}
T=T_{Born}+\bar\alpha\omega_1\omega_2\veps_1\cdot\veps_2^*+\bar\beta
(\vk_1\times\veps_1)\cdot(\vk_2\times\veps_2^*)\ , \eeq where
$\omega_i,\, \vk_i,$ and $\veps_i$ are the energy, momentum, and
polarization vector of incoming ($i=1$) and outgoing ($i=2$)
photons ($\hbar=c=1$). The contribution $T_{Born}$ corresponds
to the amplitude of Compton scattering off a  point-like
particle with spin, mass, charge, and magnetic moment equal to
those of the compound system.  For spin $S\geq 1$ the
$O(\omega^2)$ part of the Compton scattering amplitude contains
additional terms, proportional to quadrupole and higher
multipoles of the bound system \mycite{pais68}.  In particular,
for $S=1$ there is a contribution, proportional to the
quadrupole moment operator.

The investigation of electromagnetic polarizabilities is interesting
not only for systems, bound by the electromagnetic interaction, like
atoms, but also for  those, bound  by  strong interaction, such as
atomic nuclei \mycite{huett00} or hadrons \mycite{lvov93}. At present
there are many different approaches used for the description of the
electromagnetic polarizabilities of hadrons:  the MIT bag model
\mycite{hecking81,schaefer84}, the nonrelativistic quark model
\mycite{dattoli77,drechsel84,sanctis90,capstick92,liebl95}, the chiral
quark model \mycite{weiner85,scoccola90}, the chiral soliton model
\mycite{scoccola89,broniowski92} and the Skyrme model
\mycite{nyman84,chemtob87}. Here we mentioned only a small part of the
publications on these topics (see also review \mycite{lvov93a}).
Though much effort  has been devoted to these calculations, all of
them can not be considered as completely satisfactory. In particular,
there is a problem in the explanation of the magnitudes of proton and
neutron electric polarizabilities within a nonrelativistic quark
model. It was derived many years ago
\mycite{petrunkin61,baldin60,petrunkin64}
that $\bar \alpha$ can be represented as a sum
\beq \label{alpha}
\bar \alpha =
\frac23\sum_{n \neq 0} \frac{|\langle n|\vD|0 \rangle|^2}
{E_n-E_0}+ \Delta\alpha
 = \alpha_\circ + \Delta\alpha\, ,
\eeq where $\vD$ is the internal electric dipole operator,
$|0\rangle$ and $|n\rangle$ are the ground and excited states in
terms of internal coordinates, and $E_n$ and $E_0$  the
corresponding energies. The term $\Delta\alpha$ in $\bar \alpha$
has a relativistic nature  contained  and its leading term is
equal to \beqlab{deltaalpha} \Delta\alpha=\frac{e r_E^2}{3M}\, ,
\eeq where $e$ and $M$ are the particle charge and mass, $r_E$
is the electric radius defined through the Sachs form factor
$G_E$\footnote{For convenience, our definition of $r_E^2$
absorbs a total charge $e$ of the system.}.  The calculation of
the quantity $\alpha_\circ$ in the nonrelativistic quark model
without relativistic corrections taken into account leads to the
same magnitude of $\alpha_\circ$ for  proton and neutron.  Since
$\Delta\alpha$ is equal to zero for the neutron but gives a
significant contribution to $\bar\alpha$ for the proton, one has
a contradiction between the theoretical prediction of
$\bar\alpha$ for nucleons and their experimental values, since
the latter are close to each other.  In fact, this approach is
not consistent, because there are relativistic corrections to
$\alpha_\circ$ which are of the same order as $\Delta\alpha$.

Starting from  second-order perturbation theory one gets the
following expression for $\alpha_\circ$: \beqlab{alpha0}
\alpha_\circ = \frac23\sum_{n \neq 0} \frac{|\langle n|\vJ|0
\rangle|^2} {(E_n-E_0)^3} \, , \eeq where  $\vJ$ is the internal
electromagnetic current. Using the identity $\vJ=i [H,\vD]$,
where $H$ is the Hamiltonian, one  comes to the form \eq{alpha}
for  $\alpha_\circ$. The relativistic corrections to
$\alpha_\circ$  in \eq{alpha0} come from the corrections to wave
functions and energies of the ground and excited states, and
correction to current $\vJ$. Note, that in \mycite{lvov98} (see
also \mycite{bernabeu98}) it was mentioned, that such
corrections could exist, but no explicit calculations were made
for a realistic system and the importance of these corrections
was not realized.

Due to the relation between $\vJ$ and $\vD$ it is clear that there also  is a
relativistic correction to the electric  dipole moment operator (see below)  which is
connected with  the appropriate relativistic definition  of the center-of-mass
coordinate. The neglect of this relativistic correction leads to an incomplete
expression for $\bar \alpha$, and the missing  piece which is calculated in
the following turns out to be very essential.
We expect, that the inclusion of all relativistic corrections
allow one to remove the big difference
between the predictions of the nonrelativistic quark model
for proton and neutron electric polarizabilities due to the difference
in $\Delta \alpha$.

The expression for the magnetic polarizability in the nonrelativistic
quark model with no exchange and momentum-dependent forces has the
form \mycite{petrunkin64,ericson73,friar75}.
\beqlab{beta}
{\bar \beta}
= \frac23\sum_{n \neq 0} \frac{|\langle n|\vM|0 \rangle|^2}
{E_n-E_0}-\sum_i \frac{e_i^2 \langle
r_i^2\rangle}{6m_i}-\frac{\langle\vD^2\rangle}{2M}\, ,
\eeq
where $\vM$ is the internal magnetic dipole operator and the
summation in the second term on the r.h.s. is performed over the constituent
quarks, $\vr_i$ being the corresponding internal radius vector.

In order to understand the importance of the different relativistic
corrections for the polarizabilities, it is useful to consider the example
of a system where the relativistic corrections can be obtained
analytically. In this paper we calculate the low-energy Compton
scattering amplitude for  a system of two particles with
masses $m_{1,2}$ and charges $e_{1,2}$, bound by
electromagnetic forces. We consider the case $e_{1,2}^2\ll 1$ which
provides the validity of the nonrelativistic expansion. We consider
in detail the cases of spin $0$ and $1/2$ of the particles and give
the result for general case of arbitrary spins.

\section{Scattering amplitudes}

For the  electromagnetic interaction between particles  a simple
estimate shows that the part $t$ of Compton scattering
amplitude, proportional to $\omega^2$ has the form:
\beqlab{tpol} t = \omega^2 a^3
\left(c_1+c_2\frac{\varepsilon_0}{\mu}+c_3\frac{\varepsilon_0^2}{\mu^2}+\cdots\right)\
, \eeq where $a=1/(\mu g)$ is the Bohr radius, $g=-e_1 e_2 >0$,
$\mu=m_1m_2/(m_1+m_2)$ is the reduced mass, $\varepsilon_0=-\mu
g^2/2$ is the ground state binding energy in the nonrelativistic
approximation, and $c_i$ are some quantities, bilinear with
respect to $\veps_1$, $\veps_2^*$ and depending on the ratio of
charges and masses. Here, for definiteness, we assume
$\omega=\omega_1$. The two first terms of this expansion contain
the parameter $g\ll 1$ in the denominators and, therefore, come
from the contribution of big distances $r\sim a$ (or small
momenta $p\sim 1/a=\mu g$) to the matrix element. These two
terms which have no contributions from the Born amplitude are
the ones we are going to calculate in this article. Since they
are determined by a  contribution from big distances (small
momenta), it is possible to use the nonrelativistic expansion in
the calculations.  In fact, the first term is known and is
proportional to the electric polarizability $\alpha_\circ$,
calculated in the leading nonrelativistic approximation (see
below).  Some contributions to the second term are also known,
namely, those, containing the magnetic polarizability $\beta$
\eq{beta} and the correction $\Delta\alpha$ \eq{deltaalpha}.
These contributions come from the expansion of the photon wave
functions over $\vk\vr\sim \omega/\mu g$ and from the expansion
of the propagator of the system with respect to the photon
energy and the center-of-mass kinetic energy of intermediate
states.  The corresponding results for $\beta$ and
$\Delta\alpha$ can be obtained using the nonrelativistic
Hamiltonian of the system. As was mentioned above, the other
source of the contributions to $c_2$, which has not been
investigated so far, is the relativistic correction to the
Hamiltonian of the system and the corresponding corrections to
the wave functions, energy levels, and currents. We will obtain
the complete result for the second term in the expansion
\eq{tpol}. In the expression for the electromagnetic current we
neglect for a while the dependence of the form factors on the
momentum transfer. We will take this dependence into account at
the consideration of the general case of arbitrary spins.

\subsection{The system of two spin-$0$ particles}

Let us consider first the bound state of two spin-$0$ particles.
In order to calculate the Compton scattering amplitude, it is
convenient to put the system into the external
electromagnetic field $\vA(\vx,t)$. In this case the nonrelativistic
Hamiltonian has the form
\beqlab{H0A}
\tilde{H}_{nr}[\vA]=
\frac{\vpi_1^2}{2m_1}+\frac{\vpi_2^2}{2m_2}-\frac{g}{|\vr_1-\vr_2|}\,
, \eeq
where $\vpi_i=\vp_i-e_i \vA(\vr_i,t)$. Let us pass to the
variables $\vr$ and $\vR$, corresponding to the relative and
center-of-mass coordinate:
\beqlab{Rr}
\vr_1= \vR+\frac{m_2}{M}\vr\, , \quad
\vr_2= \vR-\frac{m_1}{M}\vr\, ,\quad M=m_1+m_2\, .
\eeq
Then, the momenta $\vp_i$ are
\beqlab{Pp}
\vp_1=\frac{m_1}{M}\vP +\vp\, ,
\quad
\vp_2=\frac{m_2}{M}\vP -\vp\, ,
\eeq
where $\vP=-i\grad_R$ and $\vp=-i\grad_r$.
For $\vA=0$ we have
\beqlab{H0}
\tilde{H}_{nr}[\vA=0]=\frac{\vP^2}{2M}+H_{nr}=
\frac{\vP^2}{2M}+\frac{\vp^2}{2\mu}-\frac{g}{r}\, .
\eeq
The first relativistic correction $H_B$ (Breit Hamiltonian, see,
e.g., \mycite{pilkuhn})
to \eq{H0A} reads
\beqlab{HbA}
\tilde{H}_B[\vA]= - {1\over
8}\left(\frac{\left({\vpi}_1^2\right)^2}{m_1^3}+
\frac{\left({\vpi}_2^2\right)^2}{m_2^3}\right)
+\frac{g}{2m_1m_2}\left(\frac{\delta^{ij}}{r}+
\frac{r^ir^j}{r^3}\right)\pi_1^i\pi_2^j\label{breit}\, .
\end{equation}
The first term in \eq{breit} is the correction to the
kinetic energy and the second one is the correction due to magnetic
quanta exchange, corresponding to  the space component of the photon
propagator in the Coulomb gauge. If  $\vA=0$ then  in the  center of mass
frame where the eigenvalue of the operator $\vP$ is equal to zero we
have
\beqlab{Hb}
\tilde{H}_B[\vA=0]{\Bigl|}_{\vP=0}\equiv H_B = - {1\over
8}\left(\frac{1}{m_1^3}+ \frac{1}{m_2^3}\right) \left({\bf
p}^2\right)^2-\frac{g}{2m_1m_2}\left(\frac{\delta^{ij}}{r}+
\frac{r^ir^j}{r^3}\right)p^ip^j\label{breit1},
\end{equation}
The terms, containing the operator $\vP$ in the Hamiltonian,
determine the contribution of recoil effect to the Compton scattering
amplitude. Within the precision of the present calculations these
terms should be taken into account only in the Hamiltonian
$\tilde{H}_{nr}$ and can be ommited in $\tilde{H}_B$ (see below).
The correction $\delta \varepsilon_0$ to the ground state energy, related to
the Hamiltonian $H_B$ reads
\begin{equation}
\delta \varepsilon_0=\langle 0|H_B|0\rangle = -g^4\left[{5\over 8}\mu^4
\left(\frac{1}{m_1^3}+ \frac{1}{m_2^3}\right)
+\frac{\mu^3}{m_1m_2}\right]\label{E1}.
\end{equation}
Let us start the calculation of the Compton scattering amplitude
with the amplitude $T_{nr}$ obtained with the use of the
nonrelativistic Hamiltonian \eq{H0}. This amplitude can be
represented as a sum $T_{nr}=T_{res}+T_{s}$ of resonance and
seagull parts. The part $T_{res}$ is determined by the second order
of perturbation theory with respect to the terms in
$\tilde{H}_{nr}[\vA]$, linear in the vector potential $\vA$. In the
lab frame it has the form
\beqnlab{Tres}
T_{res}&=&- \langle
\psi_0 |\exp[-i(\vk_1-\vk_2)\vR]\veps_2^*\cdot
\Biggl[\frac{e_1}{m_1}\vp_1\exp(-i\vk_2\vr_1)+
\frac{e_2}{m_2}\vp_2\exp(-i\vk_2\vr_2)\Biggr]
\times\nonumber
\\
&\times&[\varepsilon_0+\omega_1-\frac{\vP^2}{2M}-H_{nr}]^{-1}
\veps_1\cdot
\Biggl[\frac{e_1}{m_1}\vp_1\exp(i\vk_1\vr_1)+
\frac{e_2}{m_2}\vp_2\exp(i\vk_1\vr_2)\Biggr]|\psi_0\rangle+\nonumber\\
&+&(\veps_1\leftrightarrow \veps_2^*\, ,\ \omega_1\leftrightarrow
-\omega_2\, ,\ \vk_1\leftrightarrow -\vk_2 )\, .
\eeqn
Here $\psi_0(\vr)= \pi^{-1/2}(\mu g)^{3/2}\exp(-\mu g r)$ is the wave
function of the ground state, depending on the relative coordinate
$\vr$.  The final momentum of the bound system is equal to
$\vk_1-\vk_2$. Using the relations \eq{Rr} and \eq{Pp} and
making a simple transformation in order to cancel the exponents containing
$\vR$, we obtain
\beqnlab{Tres1}
T_{res}&=&- \langle \psi_0 |\veps_2^*\cdot
\Biggl[\frac{e_1}{m_1}({\bf p}+{m_1\over M}{\bf k}_1)
\exp(-i{m_2\over M}{\bf k}_2{\bf r})-
\frac{e_2}{m_2}({\bf p}-{m_2\over M}{\bf k}_1)
\exp(i{m_1\over M}{\bf k}_2{\bf r})\Biggr]\times\nonumber
\\
&\times&
G(\omega_1)\,\veps_1\cdot {\bf p}
\Biggl[\frac{e_1}{m_1}\exp(i{m_2\over M}{\bf k}_1{\bf r})-
\frac{e_2}{m_2}\exp(-i{m_1\over M}{\bf k}_1{\bf
r})\Biggr]|\psi_0\rangle+\nonumber\\
&+&(\veps_1\leftrightarrow \veps_2^*\, ,\ \omega_1\leftrightarrow
-\omega_2\, ,\ \vk_1\leftrightarrow -\vk_2 )\, .
\eeqn
Here $G(\omega)=[\varepsilon_0+\omega-\omega^2/2M-H_{nr}]^{-1}$ is the
nonrelativistic propagator of the system in the operator form.
The seagull amplitude $T_s$ is determined by  first order
of perturbation theory with respect to the terms in
$\tilde{H}_{nr}[\vA]$ which are  quadratic in $\vA$. Similar to \eq{Tres1},
we obtain
\beq \label{Tseagull}
T_{s}=-\veps_1\cdot\veps_2^*\langle\psi_0|
\left[\frac{e_1^2}{m_1}\exp(i{m_2\over M}(\vk_1-\vk_2){\bf r}) +
\frac{e_2^2}{m_2}\exp(-i{m_1\over M}(\vk_1-\vk_2){\bf r})\right]
|\psi_0\rangle\, .
\eeq
Performing the expansion of \eq{Tres} and \eq{Tseagull} with respect
to ${\bf k}_{1,2}$ and $\omega_{1,2}$ up to quadratic terms and using
the relation $\omega_1-\omega_2=(\vk_1-\vk_2)^2/2M$ , we obtain
\beqnlab{T1}
T_{nr}&=&-\veps_1\veps_2^*\frac{(e_1+e_2)^2}{M}
+\veps_1\veps_2^*\omega^2\left[\frac{9}{2\mu g^4}
\left(\frac{e_1}{m_1}-\frac{e_2}{m_2}\right)^2
+\frac{e_1+e_2}{Mg^2}\left(\frac{e_1}{m_1^2}+\frac{e_2}{m_2^2}\right)
\right]-\nonumber \\
&&-[\veps_1\times {\bf k}_1][\veps_2^*\times{\bf k}_2]
\left[\frac{1}{2g^2}\left(\frac{e_1^2}{m_1^3}+\frac{e_2^2}{m_2^3}\right)
+\frac{3}{2Mg^2}\left(\frac{e_1}{m_1}-\frac{e_2}{m_2}\right)^2
\right].
\end{eqnarray}
There is no need here to distinguish between $\omega_1$ and $\omega_2$
in the $O(\omega^2)$ term. Therefore, we set
$\omega_1=\omega_2=\omega$ in \eq{T1}. The result \eq{T1} is in agreement with
\eq{alpha} and \eq{beta}, with $\alpha_\circ$ calculated in the
nonrelativistic approximation, since in our model
\beqnlab{Dnr}
&&\vD_{nr}=\mu\left(\frac{e_1}{m_1}-\frac{e_2}{m_2}\right)\vr\,
,\nonumber\\
&&
\alpha_{\circ\, nr}= \frac23
\langle\psi_0|\vD_{nr}G_0\vD_{nr}|\psi_0\rangle =\frac{9}{2\mu g^4}
\left(\frac{e_1}{m_1}-\frac{e_2}{m_2}\right)^2,\nonumber\\
&&
\Delta\alpha=\frac{e_1+e_2}{3M}
\langle \psi_0|e_1r_1^2+e_2r_2^2|\psi_0 \rangle=
\frac{e_1+e_2}{M g^2}\left(\frac{e_1}{m_1^2}+\frac{e_2}{m_2^2}\right)\, , \\
&&
\langle \psi_0|\vM|\psi_{n\neq 0}\rangle=0\, , \quad \langle
\psi_0|{\bf D}^2|\psi_0 \rangle=
\frac{3}{g^2}\left(\frac{e_1}{m_1}-\frac{e_2}{m_2}\right)^2\,
,\nonumber\\
&& \langle \psi_0|\left[
\frac{e_1^2 r_1^2}{m_1}+\frac{e_2^2
r_2^2}{m_2}\right]|\psi_0\rangle=
\frac{3}{g^2}\left(\frac{e_1^2}{m_1^3}+\frac{e_2^2}{m_2^3}\right)\, .
\nonumber
\end{eqnarray}
Here $G_0$ is the reduced Green function in the operator form:
\beqlab{G0}
G_0=[\varepsilon_0-H_{nr}+i0]^{-1}
(1-|\psi_0\rangle\langle\psi_0|)\, .
\eeq
The details of calculations of different matrix elements, containing the
operator $G_0$ are presented in Appendix A.

We pass now to the calculation of the relativistic corrections to
the electromagnetic polarizabilities, connected with the Breit
Hamiltonian $\tilde{H}_B$ \eq{HbA}. We perform the calculations in
the same way as at the derivation of \eq{Tres} and \eq{Tseagull}, but
for the Hamiltonian $\tilde{H}=\tilde{H}_{nr}+\tilde{H}_B$. At the
calculation of these corrections within our accuracy the terms of the
Hamiltonian $\tilde{H}_B[\vA]$, quadratic in $\vA$ do not contribute
to the electromagnetic polarizabilities, i.e. the seagull
contribution from $\tilde{H}_B[\vA]$ is absent. In the corrections to the
electromagnetic polarizabilities from the resonance part of the
amplitude, we can take the second order of expansion with respect to
$\omega$ of the operator Green function and put $\vk_{1,2}=0$
elsewhere. This  means that within our accuracy we can neglect in
$\tilde{H}_B[\vA]$ the terms  containing the total momentum  $\vP$ and
replace the exponents in the photon wave function by unity.
Therefore, the terms in the Hamiltonian $\tilde{H}_B[\vA]$, linear in
$\vA$, can be represented in the form $-\vA(0)\vJ_B$, where
\beqlab{Jb}
\vJ_B=-\left(\frac{e_1}{m_1^3}-\frac{e_2}{m_2^3}\right)\frac{\vp^2}{2}\vp
-\frac{g(e_1-e_2)}{2m_1m_2
r}\left(\vp+\frac{\vr}{r^2}(\vr\cdot\vp)\right)
\eeq
is the correction to the operator of the total internal current $\vJ$ :
\beqlab{Jnb}
\vJ=\vJ_{nr}+ \vJ_B= (e_1/m_1-e_2/m_2)\vp +\vJ_B \, .
\eeq

Let us now discuss  the relativistic correction to the electric dipole moment
 operator.
In the lab frame it is equal to $\vD_{tot}=e_1\vr_1+e_2\vr_2$.
For the total Hamiltonian of the system the following relation holds
\beqnlab{Jtot}
&&
i\Bigl[\tilde{H}_{nr}[\vA=0]+\tilde{H}_B[\vA=0],\, \vD_{tot} \Bigr]=
\\
&&
=e_1\frac{\vp_1}{m_1}\left(1-\frac{\vp_1^2}{2m_1^2}\right)
-\frac{g\, e_1}{2m_1m_2r}
\left(\vp_1+\frac{\vr}{r^2}(\vr\cdot\vp_1)\right)+
(1\leftrightarrow 2)\, .\nonumber
\eeqn
The r.h.s. of \eq{Jtot} is nothing  but the operator of total
current in the lab frame, containing the total momentum $\vP$.
This can be verified
by differentiating the Hamiltonian $\tilde{H}_{nr}[\vA]+\tilde{H}_B[\vA]$
over $\vA$ at $\vA=0$. Therefore, there are no  relativistic
corrections to the total electric dipole operator $\vD_{tot}$.
The internal electric dipole moment operator is defined as
\beqlab{vd}
\vD=\vD_{tot}-(e_1+e_2)\vR_{cm}\ ,
\eeq
where $\vR_{cm}$ is the center-of-mass vector.
This vector is defind in such a way that it satisfies the following relations:
\beqlab{rcmrel}
[\vR_{cm}\, ,\, \vP]=i \quad ,\quad
i[H_{tot}\, ,\, \vR_{cm}]=\frac{\vP}{H_{tot}} \ ,
\eeq
where $H_{tot}$ is the total relativistic Hamiltonian of the system, and $\vP$
 is the  total momentum. Within
 our accuracy the second relation in \eq{rcmrel} reads:
\beqlab{rcmrel1}
i[\tilde H_{nr}[\vA=0]+ \tilde H_{B}[\vA=0]\, ,\,\vR_{cm} ]=
\frac{\vP}{M}\left(1-\frac{\tilde H_{nr}[\vA=0]}{M}\right) \, .
\eeq
 It is known (see, e.g. , \mycite{landau}) that there is a  relativistic
correction to  $\vR_{cm}$ in classical electrodynamics. For the
case of two particles, interacting due to electromagnetic field,
the corresponding operator which satisfies the relations
\eq{rcmrel} has the form \beqlab{Rcm} \vR_{cm}=\vR+
\frac{1}{2M}\left(\left\{\vr_1\, ,\,
\frac{\vp_1^2}{2m_1}-\frac{g}{2r}\right\}+ \left\{\vr_2\, ,\,
\frac{\vp_2^2}{2m_2}-\frac{g}{2r}\right\}- \left\{\vR\, ,\,
\frac{\vp_1^2}{2m_1}+ \frac{\vp_2^2}{2m_2} -
\frac{g}{r}\right\}\right) \, . \eeq Here we took into account
the first relativistic correction and use the  notation
$\{a,b\}=ab+ba$. In terms of the variables $\vr$ and $\vp$ (see
\eq{Rr} and \eq{Pp}) we obtain : \beqlab{Rcm1}
\vR_{cm}=\vR+\frac{(m_2-m_1)}{2M^2}\left(\left\{\vr ,\,
H_{nr}\right\}+ g\frac{\vr}{r}\right) \, , \eeq where the term
proportional to the total momentum $\vP$ is omitted.
Substituting this expression into \eq{vd} we obtain the
relativistic correction to the internal electric dipole moment:
\beqlab{dbnew}
\vD_{B}=\frac{(e_1+e_2)(m_1-m_2)}{2M^2}\left(\left\{\vr, \,
H_{nr}\right\}+ g\frac{\vr}{r}\right) \, . \eeq Note that, as
should be the case, the operator of total internal current
$\vJ=\vJ_{nr}+\vJ_B$ satisfies within our accuracy  the relation
\beqlab{jdrel} \vJ=i[H_{nr}+H_B,\vD_{nr}+\vD_B]\, , \eeq
 where $\vD_{nr}$ is defined in \eq{Dnr}, and $H_{nr}+H_B$ is the
 internal part of Hamiltonian (see \eq{H0} and \eq{Hb}).

Let us write down now the corrections to the $O(\omega^2)$ term of
Compton scattering amplitude, related to the Breit Hamiltonian and
the corresponding current.
The correction due to $\vJ_B$ reads
\beqlab{mc}
t_{c}=-\omega^2
\langle \psi_0 |[
\veps_2^*\cdot \vJ_B\, G_0^3\,\veps_1\cdot\vJ_{nr} +
\veps_2^*\cdot \vJ_{nr}\, G_0^3\,\veps_1\cdot\vJ_B
]|\psi_0\rangle +(\veps_1\leftrightarrow \veps_2^*)\, .
\end{equation}
The $O(\omega^2)$ correction to the amplitude, connected with the
expansion of the propagator with respect to $H_B$, has the form
\beqlab{mp}
t_{p}=
-\omega^2
\langle \psi_0 |
\veps_2^*\cdot {\vJ_{nr}}[G_0^2H_BG_0^2+G_0^3H_BG_0+G_0H_BG_0^3]
\veps_1\cdot {\vJ_{nr}}
|\psi_0\rangle +(\veps_1\leftrightarrow \veps_2^*)\, .\nonumber
\eeq
The contribution due to the correction to wave function is
\beqlab{mw}
t_{w}= -\omega^2 \langle \psi_0 |[
\veps_2^*\cdot {\vJ_{nr}}G_0^3\veps_1\cdot {\vJ_{nr}}
G_0 H_B+  H_B G_0 \veps_2^*\cdot {\vJ_{nr}}G_0^3\veps_1\cdot {\vJ_{nr}}
]|\psi_0\rangle +(\veps_1\leftrightarrow \veps_2^*) \, .\nonumber
\eeq
At last, the contribution  corresponding to the correction to the
ground state energy  reads:
\beqlab{me}
t_{e}= 3\omega^2\delta \varepsilon_0
\langle \psi_0 |
\veps_2^*\cdot {\vJ_{nr}}G_0^4\veps_1\cdot {\vJ_{nr}}
|\psi_0\rangle +(\veps_1\leftrightarrow \veps_2^*) .\nonumber
\end{equation}
In order to calculate the matrix elements in \eq{mc}-\eq{me} it is
convenient to use the following relations (see Appendix A):
\beqnlab{rel1}
&&G_0 \, {\bf r}\psi_0=-\frac{{\bf
r}}{g}\left(\frac{r}{2}+a\right)\psi_0, \quad G_0 \, r{\bf
r}\psi_0=-\frac{{\bf r}}{g}\left(\frac{r^2}{3}+
\frac{5ar}{6}+\frac{5a^2}{3}\right)\psi_0, \nonumber \\
&&
G_0 r^2\psi_0=-\frac{1}{g}\left(\frac{r^3}{3}+
ar^3-\frac{11a^3}{2}\right)\psi_0 ,\\
&&
G_0 r^3\psi_0=-\frac{1}{g}\left(\frac{r^4}{4}+
\frac{5ar^3}{6}+\frac{5a^2r^2}{2}-\frac{155a^4}{8}\right)\psi_0
,\nonumber
\end{eqnarray}
where $a=1/\mu g$.  Using (\ref{rel1}) and
also the relation ${\bf p}=i\mu\, [H_{nr},{\bf r}]$ , we get the
following results for the contributions \eq{mc}-\eq{me}
\beqnlab{ta}
t_c&=&-\frac{2\mu\omega^2 \veps_1\cdot\veps_2^*}{9g^2}
\left(\frac{e_1}{m_1}-\frac{e_2}{m_2}\right)
\left[20\left(\frac{e_1}{m_1^3}-\frac{e_2}{m_2^3}\right)
+\frac{89}{4}\frac{(e_1-e_2)}{\mu m_1m_2}\right]\, ,\\
t_p&=&\frac{\mu^2\omega^2 \veps_1\cdot\veps_2^*}{g^2}
\left(\frac{e_1}{m_1}-\frac{e_2}{m_2}\right)^2\left[
\frac{1061}{288}\left(\frac{1}{m_1^3}+\frac{1}{m_2^3}\right)+
\frac{25}{3\mu m_1m_2}\right]\nonumber\, ,\\
t_w&=&\frac{\mu^2\omega^2 \veps_1\cdot\veps_2^*}{g^2}
\left(\frac{e_1}{m_1}-\frac{e_2}{m_2}\right)^2\left[
\frac{3}{4}\left(\frac{1}{m_1^3}+\frac{1}{m_2^3}\right)+
\frac{14}{9\mu m_1m_2}\right]\nonumber\, ,\\
t_e&=&-\frac{\mu^2\omega^2 \veps_1\cdot\veps_2^*}{g^2}
\left(\frac{e_1}{m_1}-\frac{e_2}{m_2}\right)^2\frac{129}{4}\left[
\frac{5}{8}\left(\frac{1}{m_1^3}+\frac{1}{m_2^3}\right)+
\frac{1}{\mu m_1m_2}\right]\, .\nonumber
\eeqn
Representing the sum of all contributions in \eq{ta} as $\omega^2
\veps_1\cdot\veps_2^* \alpha_{\circ\, B}$ we obtain
the following result for $\alpha_{\circ\, B}$ :
\beqlab{alphab}
\alpha_{\circ B}=-\frac{1}{g^2}
\left(\frac{e_1}{m_1}-\frac{e_2}{m_2}\right)^2
\left(\frac{121}{6\mu}-\frac{113}{4M}\right)
-\frac{(e_1+e_2)}{Mg^2}\left(\frac{e_1}{m_1}-\frac{e_2}{m_2}\right)
\frac{(m_1-m_2)}{2 m_1m_2} \, .
\eeq
If one starts the calculation of  $\alpha_{\circ}$
from \eq{alpha}, then it is easy to check that the first term in
\eq{alphab} corresponds to the sum of   corrections due to modification
of wave function, propagator, and ground state energy. The second term in
\eq{alphab} corresponds to the contribution due to the relativistic correction
to the electric dipole moment \eq{dbnew}. Therefore, this correction
 appears due to the correct description of  the center-of-mass motion. It is
seen that the first term in \eq{alphab} has the same dependence on
charges as  $\alpha_{\circ nr}$   while the second term is
similar to $\Delta\alpha$ , \eq{Dnr}.
Taking a sum of $\alpha_{\circ B}$ , $\alpha_{\circ\, nr}$, and
$\Delta\alpha$, we come to the following result for
$\bar\alpha$ for the system under consideration:
\beqnlab{alphafinal}
\bar\alpha&=&\frac{1}{\mu g^4}
\left(\frac{e_1}{m_1}-\frac{e_2}{m_2}\right)^2
\left[\frac{9}{2}-g^2\left(\frac{121}{6}-\frac{113\mu}{4M}\right)\right]+
\nonumber\\
&&+\frac{(e_1+e_2)}{Mg^2}\left[\frac{3}{2}\left(\frac{e_1}{m_1^2}+
\frac{e_2}{m_2^2}\right)-\frac{(e_1+e_2)}{2 m_1m_2}\right] \, .
\eeqn
Thus, the relativistic corrections  to $\alpha_{\circ}$ has
reduced to a  renormalization of $\alpha_{\circ\, nr}$ and essentially to a
 modification of
$\Delta \alpha$ , \eq{deltaalpha}. One can expect that the last statement
is valid not only for the system under consideration. Indeed,
due to the definition of $\vD$,  the correction to
$\bar\alpha$ related to the modification of $\vR_{cm}$  is
proportional to $Q/M$ , where Q is the  total charge of the system, and,
therefore, has the same structure as $\Delta  \alpha$.

As a nontrivial test of our method of calculation we checked the
fulfilment of the low-energy theorem for the Compton scattering
amplitude. At $\omega=0$ this amplitude should have the form
\beqlab{T0}
T(\omega=0)=-\veps_1\cdot\veps_2^*\frac{(e_1+e_2)^2}{E_0}\approx
-\veps_1\cdot\veps_2^*\frac{(e_1+e_2)^2}{M+\varepsilon_0}\approx
-\veps_1\cdot\veps_2^*\frac{(e_1+e_2)^2}{M}
\left(1-\frac{\varepsilon_0}{M}\right),
\eeq
where $E_0=M+\varepsilon_0$ is the mass of the system.
It is interesting, that the term in r.h.s. of \eq{T0}, proportional to the
nonrelativistic energy $\varepsilon_0$, appears as a contribution of terms
from the Breit Hamiltonian $\tilde{H}_B[\vA]$, which we checked by explicit
calculations (see Appendix B).

\subsection{The system of a spin-$0$ particle and a spin-$1/2$ particle}

Let the first particle have the spin $1/2$ and the second particle
have the spin $0$.
Then we should add the term
\beqlab{pauli}
\delta\tilde{H}_{nr}[\vA]=
-\frac{e_1(1+\kappa_1)}{m_1}\vs_1\cdot\vH
\eeq
to the nonrelativistic Hamiltonian $\tilde{H}_{nr}[\vA]$, \eq{H0A}.
Here $\vH$ is the external magnetic field, $\vs_1=\vsig_1/2$ is the
spin operator of the first particle, and $\kappa_1$ is its
anomalous magnetic moment in units $e_1/2m_1$. There is  also some
additional contribution $\delta\tilde{H}_B[\vA]$ to
$\tilde{H}_B[\vA]$, \eq{Hb} (see, e.g. \mycite{pilkuhn}). The  terms
of $\delta\tilde{H}_B[\vA]$ linear in
$\vs_1$ as well as \eq{pauli}
determine the $O(\omega)$ terms of the Compton amplitude. These terms are
well-known and follow, together with the $\omega$-independent
term, from the low-energy theorem \mycite{low}.
As it was explained in the previous subsection, it is
sufficient within our accuracy to account for  the Breit Hamiltonian
only in the long-wave
limit, i.e. at $\omega_{1,2}=0$ in order to obtain the $O(\omega^2)$
terms of Compton amplitude. In this limit the Hamiltonian
$\delta\tilde{H}_B[\vA]$ reads
\beqlab{dhb}
\delta\tilde{H}_B[\vA]=-\frac{e_1e_2(1+2\kappa_1)}{2m_1^2}\left(
\pi\delta(\vr)+\frac1{r^3}\vs_1\cdot(\vr\times\vpi_1)
\right)+\frac{e_1e_2(1+\kappa_1)}{m_1m_2r^3}
\vs_1\cdot(\vr\times\vpi_2) \, .
\eeq
The explicit calculation shows that the contribution  of
$\delta\tilde{H}_{nr}[\vA]$ given by \eq{pauli} and the  terms
in \eq{dhb} linear in $\sigma_1$ do not lead to any contributions
to $c_{1,2}$ in
\eq{tpol}, i.e. they can be neglected in  the calculation of
polarizabilities within our accuracy. In particular, there are no terms
linear in the spin  in the quantities $c_{1,2}$, which is in
agreement with the general conclusion on the absence of terms
$O(\omega^2)$ linear in spin  in the non-Born part of the Compton
amplitude \mycite{lvov98}. The only term which should be
taken into account in addition to those considered in the previous
subsection, is the spin-independent term in \eq{dhb} (Darwin term):
\beqlab{darvin}
\delta_D H_B=\frac{\pi g(1+2\kappa_1)}{2m_1^2}\delta(\vr)\, .
\eeq
It follows from \eq{dhb} that there is no correction to the current
associated with the Hamiltonian $\delta_D H_B$. Using the expressions
(\ref{mp}-\ref{me}) with the replacement $H_B\to\delta_D H_B$ and the
relations \eq{rel1} we obtain
\beqn
&&
\delta_D t_c=0\, ,\quad\delta_D t_w=-\frac{5\mu\omega^2
\veps_1\cdot\veps_2^*}{8m_1^2g^2}
\left(\frac{e_1}{m_1}-\frac{e_2}{m_2}\right)^2(1+2\kappa_1) \, ,\\
&&
 \delta_D t_p=0\, ,\quad\delta_D t_e=\frac{129\mu\omega^2
 \veps_1\cdot\veps_2^*}{8m_1^2g^2}
\left(\frac{e_1}{m_1}-\frac{e_2}{m_2}\right)^2(1+2\kappa_1)
\, .\nonumber
\eeqn
As a result, the correction to the electric polarizability
associated with the Breit Hamiltonian in the system of spin-$0$ and
spin-$1/2$ will be the sum of $\alpha_{\circ\, B}$ \eq{alphab} and
\beqlab{d1alphab}
\delta \alpha_{\circ B}=\frac{31\mu}{2m_1^2g^2}
\left(\frac{e_1}{m_1}-\frac{e_2}{m_2}\right)^2(1+2\kappa_1)\, .
\eeq

\subsection{The system of two spin-$1/2$ particles}

In the case of two spin-$1/2$ particles it is necessary to account for
two Darvin terms  in addition to the Breit
Hamiltonian \eq{Hb}, corresponding to both particles
\beqlab{darvinp}
\delta_D H_B=\frac{\pi g(1+2\kappa_1)}{2m_1^2}\delta(\vr)+
\frac{\pi g(1+2\kappa_2)}{2m_2^2}\delta(\vr)
\eeq
and the Hamiltonian, corresponding to spin-spin interaction
\mycite{pilkuhn}:
\beqlab{spinspin}
\delta_s H_B=\frac{g(1+\kappa_1)(1+\kappa_2)}{m_1m_2}
\left[\frac{3(\vn\cdot\vs_1)(\vn\cdot\vs_2)-\vs_1\cdot\vs_2}{r^3}
+\frac{8\pi}3\delta(\vr)\, \vs_1\cdot\vs_2\right]\, ,
\eeq
with $\vn=\vr/r$. It is more convenient to rewrite $\delta_s H_B$ in
terms of the total spin operator $\vS=\vs_1+\vs_2$:
\beqlab{spinspin1}
\delta_s H_B=\frac{g(1+\kappa_1)(1+\kappa_2)}{2m_1m_2}
\left[\frac{3n_in_j Q_{ij}}{2r^3}+
4\pi\delta(\vr)\, \left(\frac23\vS^2-1\right)\right]\, ,
\eeq
where the operator $Q_{ij}$, quadratic in $\vS$, is equal to
\beqlab{Qij}
Q_{ij}=S_iS_j+S_jS_i-\frac23\vS^2\delta_{ij}\, .
\eeq
Note that in such a system as positronium it is necessary to add the
contribution of the annihilation diagram, which results in the
replacement $(2\vS^2/3-1)\to (7\vS^2/6-1)$ in the coefficient of the
$\delta$-function in \eq{spinspin1} (of course, in this case
$m_1=m_2$, $e_1=-e_2$).
As in the previous subsection, the terms proportional to the
$\delta$-function in \eq{darvinp} and \eq{spinspin1} give the
contribution $\delta\alpha_{\circ B}$, which should be added to
$\alpha_{\circ_B}$, \eq{alphab}:
\beqlab{d2alphab}
\delta \alpha_{\circ B}=
\frac{31\mu}{2g^2}
\left(\frac{e_1}{m_1}-\frac{e_2}{m_2}\right)^2
\left[\frac{1+2\kappa_1}{m_1^2}+\frac{1+2\kappa_2}{m_2^2}+
\frac{4(1+\kappa_1)(1+\kappa_2)}{m_1m_2}\left(\frac23S(S+1)-1\right)
\right]\, .
\eeq
Here we replaced $\vS^2$ by its eigenvalue $S(S+1)$, where $S=0,1$ is
the total spin of the system.
The term in \eq{d2alphab} containing the tensor operator
$3(\vn\cdot\vS)^2-\vS^2$ determines the contribution to the
$O(\omega^2)$ part of the Compton amplitude, which has the form
\beq
t_{(tensor)}=\omega^2\alpha_T\epsilon_1^i\epsilon_2^{j*}
\langle Q_{ij}\rangle\, ,
\eeq
where $\langle\cdots\rangle$ denotes the averaging over the spin part
of the wave function. Of course, $t_{(tensor)}$ vanishes if $S=0$.
Since there is no correction to the current or  to the energy of the
ground state due to the tensor part of $\delta_s H_B$, the
contributions to $\alpha_T$ come only from the corrections to the
propagator and to the wave function. Using \eq{mp}, \eq{mw}, and the
relations (see Appendix A)
\beqn
G_0 (3r_ir_j-r^2\delta_{ij})\ket&=&
-\frac{3r_ir_j-r^2\delta_{ij}}{g}
\left(\frac r3+\frac a2\right)\ket\\
G_0 (3r_ir_j-r^2\delta_{ij}) r\ket&=&
-\frac{3r_ir_j-r^2\delta_{ij}}{g}\left( \frac{7a^2}8 + \frac{7a
r}{12} + \frac{r^2}4 \right)\ket
\nonumber\, ,
\eeqn
we obtain
\beqlab{alphat}
\alpha_T=-\frac{47(1+\kappa_1)(1+\kappa_2)}{40Mg^2}
\left(\frac{e_1}{m_1}-\frac{e_2}{m_2}\right)^2.
\eeq

In the system of two spin-1/2 particles there is a big  paramagnetic
contribution to the magnetic polarizability from the first term in
\eq{beta}. The main contribution corresponds to the transition from the ground
state with the total spin $S=0$ to the state with $S=1$, with both states
having the same angular momenta, $l=0$, and
 radial quantum nambers $n_r=0$ (hyperfine splitting).
 Representing the spin part of the magnetic moment operator in the form
$$\vM_s=f_1\vs_1+f_2\vs_2\quad ,\quad f_i=\frac{e_i(1+\kappa_i)}{m_i}\, ,$$
and using \eq{spinspin}, we obtain:
\beqlab{betapara}
\beta_{1}=-\frac{3(f_1-f_2)^2a^3}{16f_1f_2}\quad ,
\quad a=\frac{1}{\mu g}\, .
\eeq
As was pointed out in the  previous section, for positronium it is necessary
to change the coefficient of $\delta$-function in \eq{spinspin1}.
As a result, the contribution of the first term in \eq{beta} to the
magnetic polarizability of positronium is:
\beqlab{betapos}
\beta_{1}=\pm \frac{3}{7}a^3\quad ,
\eeq
where upper sign corresponds to  parapositronium ($S=0$), and lower sign to
orthopositronium ($S=1$).

\subsection{The system of two particles with arbitrary spins}

Let the particles have the spins $s_{1,2}$ and magnetic
moments $\mu_{1,2}$ which we represent in the form
\beq
\mu_a=\frac{e_as_a}{m_a}(1+\kappa_a)\, , \quad a=1,2\,  .
\eeq
The electromagnetic current for each particle has
the form (see, e.g., \mycite{khri96,khri97})
\beqlab{formfactor}
j_\mu= \bar{\psi}(p')\left[
F_e \frac{p_\nu+p'_\nu}{2m}+
\frac{G_m}{2m}\Sigma_{\mu\,\nu}
q^\nu
\right]\psi(p)\, ,
\eeq
where $q=p'-p$. The operator  $\Sigma_{\mu\,\nu}$ is a generalisation of the
corresponding matrix for spin $1/2$. The indices
numerating the particles have been omitted.  The quantities $F_e$
and $G_m$ depend on
$q^2$ and $(s_\mu q^\mu)^2$, where $s_\mu$ is the 4-vector of the
spin operator. This quantities are normalized as follows:
\beq
F_e(q=0)=1\, ,\quad G_m(q=0)=1+\kappa\, .
\eeq
If we neglect the $q$-dependence of the form factors, then, in
addition to the Breit Hamiltonian for two spin-$0$ particles,
\eq{HbA}, it is necessary to take into account the Hamiltonian
\eq{spinspin} (with the corresponding spin operators) and two other
contributions \mycite{khri96}. Namely, the Darwin Hamiltonian
\beq
\delta_D H_B=\sum_{a=1,2}
\frac{2\pi g}{3 m_a^2}(1+2\kappa_a)
(s_a+\zeta_a)\delta(\vr) \, ,
\eeq
$\zeta=0$ for integer spin and $\zeta=1/4$ otherwise, and
the term  containing the quadrupole moments of the particles:
\beq
\delta_Q H_B=\sum_{a=1,2}
\frac{g(1+2\kappa_a)\xi_a}{2m_a^2 r^3}
(3(\vn\cdot\vs_a)^2-\vs_a^2)\, ,
\eeq
$\xi=1/(2s-1)$ for integer spin and $\xi=1/(2s)$ otherwise.
It is clear that all matrix elements can be calculated in the same
way as in the previous subsection. The averaging over the spin
variables can be done  using the following relations
\beqn
\langle S\,S_z|\left[
s_{1i}s_{2j}+s_{2i}s_{1j}-\frac23 \delta_{ij} \vs_1\vs_2
\right]|S,S_z^\prime\rangle\!\!&=&\!\!
A(S,s_1,s_2)\langle S\,S_z|Q_{ij}|S,S_z^\prime\rangle \, ,\\
\langle S\,S_z|\left[
s_{1i}s_{1j}+s_{1i}s_{1j}-\frac23 \delta_{ij} \vs_1^2
\right]|S,S_z^\prime\rangle\!\!&=&\!\!
B(S,s_1,s_2)\langle S\,S_z|Q_{ij}|S,S_z^\prime\rangle\nonumber
\eeqn
where $\vS=\vs_1+\vs_2$ is the total spin operator, $Q_{ij}$ is
defined in \eq{Qij}, and for $S\geq 1$
\beqn
A(S,s_1,s_2)&=&
\frac{\Lambda^2+2\Lambda(\lambda_1+\lambda_2)-3(\lambda_1-\lambda_2)^2}
{2\Lambda(4\Lambda-3)}\\
B(S,s_1,s_2)&=&\frac{3\Lambda^2+\Lambda(2\lambda_1-6\lambda_2-3)
+3(\lambda_1-\lambda_2)(\lambda_1-\lambda_2-1)}
{2\Lambda(4\Lambda-3)}  \nonumber\, .
\eeqn
Here $\Lambda=S(S+1)$, $\lambda_{1,2}=s_{1,2}(s_{1,2}+1)$
are the eigenvalues of the operators $\vS^2$ and $\vs_{1,2}^2$,
respectively.
For $S=0,\,1/2$ we put $A=B=0$.  As a result, we
obtain the following generalization of \eq{d2alphab} to the case of
arbitrary spins:
\beqnlab{d2alphabg}
\delta \alpha_{\circ
B}&=& \frac{62\mu}{3g^2}
\left(\frac{e_1}{m_1}-\frac{e_2}{m_2}\right)^2
\left[\frac{1+2\kappa_1}{m_1^2}(s_1+\zeta_1)+\frac{1+2\kappa_2}{m_2^2}
(s_2+\zeta_2)+\right.\\
&&
+\left.\frac{2(1+\kappa_1)(1+\kappa_2)}{m_1m_2}\left(\Lambda-\lambda_1-\lambda_2
\right)
\right]\, .\nonumber
\eeqn
The generalization of \eq{alphat} is
\beqnlab{alphatg}
\alpha_T&=&-\frac{47\mu}{40g^2}
\left(\frac{e_1}{m_1}-\frac{e_2}{m_2}\right)^2
\left[
\frac{2(1+\kappa_1)(1+\kappa_2)}{m_1m_2} A(S,s_1,s_2)+\right.\\
&&+\left.
\frac{1+2\kappa_1}{m_1^2}\xi_1B(S,s_1,s_2)
+\frac{1+2\kappa_2}{m_2^2}\xi_2B(S,s_2,s_1)
\right]\, . \nonumber
\eeqn
Let us now take into account the $q$-dependence of the
electromagnetic form factors of the constituents  defined in
\eq{formfactor}.
We assume, that the scale of variation of these form factors are much
larger than the typical momentum transfer $\sim\mu g$. In other words,
the characteristic size of each constituent is much smaller than the
size of the whole system $a=1/\mu g$. In this case it is
sufficient, within our accuracy, to take $G_m=1+\kappa$ and  to expand
the form factor $F_e$ up to quadratic in $q$ terms:
\beq
F_e(q^2,(s^\mu q_\mu)^2)\approx 1-\frac{r_e^2 \vq^2}6 +
\frac{r_s^2(\vs\cdot\vq)^2}2\, ,
\eeq
where $r_{e,s}^2$ are some constants. Multiplying the $O(\vq^2)$
terms in this expression by $-4\pi g/\vq^2$ and performing the
Fourier transform, we obtain the additional terms in the Hamiltonian
\beq
\delta_{f}H_B=\sum_{a=1,2}\left[
\frac{2\pi g}3 \left(
r_{ea}^2-r_{sa}^2\vs_a^2
\right)\delta(\vr)+
g r_{sa}^2\frac{3(\vs_a\cdot\vn)^2-\vs_a^2}{2r^3}\right]\, .
\eeq
Since the terms in this Hamiltonian have the same structure as above,
it is easy to write down the result for the corresponding corrections
to polarizabilities:
\beqnlab{ff}
\delta_{f} \alpha_{\circ B}&=&
\frac{62\mu}{3g^2}
\left(\frac{e_1}{m_1}-\frac{e_2}{m_2}\right)^2
(r_{e1}^2-r_{s1}^2\lambda_1+r_{e2}^2-r_{s2}^2\lambda_2)\, ,\\
\delta_{f} \alpha_{T}&=&
-\frac{47\mu}{40g^2}
\left(\frac{e_1}{m_1}-\frac{e_2}{m_2}\right)^2
\left[
r_{s1}^2 B(S,s_1,s_2)+r_{s2}^2 B(S,s_2,s_1) \nonumber
\right] \, .
\eeqn
If the parameters of the form factors $r_{e,s}^2\sim 1/m^2\ll a^2$, then
the contributions \eq{ff} to the polarizabilities are of the same order
as $\alpha_{\circ B}$.
The first relativistic correction to the Compton scattering
amplitude at $\omega=0$, \eq{T0}, is proportional to
$\varepsilon_0=-\mu g^2/2$ and is independent of the spins of the
constituents. Then, the correction to the amplitude at $\omega=0$
connected with spin-dependent terms in Breit Hamiltonian as well as the
Darwin terms (also having the spin origin) should vanish. This
statement was checked explicitly (see Appendix B).

Let us consider now the  paramagnetic
contribution to the magnetic polarizability from the first term in
\eq{beta}. Let $s_1\ge s_2$. Then, the total spin of the ground state is
$S=s_1-s_2$, and the main contribution corresponds to the transition
from the ground state  to the state with $S=s_1-s_2+1$, with both states
having the same angular momentum, $l=0$, and
 radial quantum namber $n_r=0$ (hyperfine splitting).
A simple explicit calculation leads to:
\beqlab{betaparageneral}
\beta_{1}=-\frac{(f_1-f_2)^2s_2(s_1+1)a^3}{4f_1f_2(s_1-s_2+1)^2}\quad .
\eeq
This term should be added to the diamagnetic contribution $\beta_{dia}$
(see \eq{T1}) :
\beqlab{dia}
\beta_{dia}=
-\frac{1}{2g^2}\left(\frac{e_1^2}{m_1^3}+\frac{e_2^2}{m_2^3}\right)
-\frac{3}{2Mg^2}\left(\frac{e_1}{m_1}-\frac{e_2}{m_2}\right)^2 \, .
\eeq

\section{Conclusion}

We have obtained the complete result for the first relativistic
corrections to the electromagnetic polarizabilities, including the tensor
part which exists for the total spin $S\geq1$. We demonstrated
that, within our accuracy, this tensor part contains the quadrupole
moment of the system and no any higher multipoles. For the system of
two spinless particles it is easy to check that the total relativistic
correction  Eqs. (\ref{Dnr}) and
(\ref{alphab}) is negative at arbitrary masses and charges.
In the general case of non-zero spins and arbitrary anomalous
magnetic moments the relativistic correction
$\Delta\alpha+\alpha_{\circ B}+\delta\alpha_{\circ B}$, where
$\delta\alpha_{\circ B}$ is given by \eq{d2alphabg}, can be
positive.
It is interesting to consider some special cases.
The first of them is a hydrogen-like ion. In this case $e_1=e$,
$e_2=-Ze$, and $m_2\gg m_1$. In the limit $m_2\to
\infty$ the result for electric polarizabilities is independent of the spin
and magnetic moment of the nucleus. Neglecting also the anomalous
magnetic moment $\kappa_1$  of the electron, we obtain from
\eq{T1}, \eq{Dnr}, \eq{alphab}, and \eq{d2alphab}
\beqlab{had}
\bar{\alpha}=\frac9{2m^3\alpha_{em}^3Z^4}
-\frac{14}{3m^3\alpha_{em} Z^2}\quad ,
\eeq
where $\alpha_{em}=e^2=1/137$ is the fine-structure constant. Note
that in this limit the correction $\Delta\alpha$ , \eq{deltaalpha},
vanishes. The result \eq{had} is in agreement with that obtained
in \mycite{manakov} with the use of the reduced  Green function
of the Dirac equation for an electron in a Coulomb field.
For the magnetic polarizability at $s_2=0$ we have
\beqlab{hadbeta}
\bar{\beta}=\beta_{dia}=-\frac{1}{2m^3\alpha_{em} Z^2}\, .
\eeq
For $s_2=1/2$ in the limit $m_2\gg m_1$ there  is
a very big contribution from the paramagnetic part of the
magnetic polarizability,
\eq{betapara}. In the Compton scattering amplitude  this
contribution should be taken into account only for
photon energies  $\omega$ much smaller than the energy
$E_{hf}\sim \alpha_{em}^4m_1^2/m_2$ of
the  hyperfine splitting. For $\mu g^2\gg\omega\gg E_{hf}$ the paramagnetic
contribution should be omitted.

Another interesting example is positronium. As we mentioned above,
in this case it is necessary to replace $(2\vS^2/3-1)\to
(7\vS^2/6-1)$ in the coefficient of the  $\delta$-function in
\eq{spinspin1} due to the contribution of the annihilation diagram.
Putting $m_1=m_2=m$, $e_1=-e_2=e$, and $\kappa_1=\kappa_2=0$, we
obtain $\Delta\alpha=0$ and the complete result for the polarizabilities
\beqn
\bar\alpha&=&\frac{36}{(m\alpha_{em})^3}+\frac1{6
m^3\alpha_{em}}\times\left\{
-1001 \quad\mbox{for } S=0
\atop
735   \quad\mbox{for } S=1
\right.\nonumber\\
\bar\beta&=&(-1)^S\frac3{56m^3\alpha_{em}^3}\,
-\frac4{m^3\alpha_{em}}\, .
\eeqn
As in the previous case, for photon energy $\omega\gg\alpha_{em}^4m$
the paramagnetic contribution should be omitted in the Compton scattering
amplitude.

For $S=1$ (orthopositronium) we also have the tensor polarizability
\beq
\alpha_T=-\frac{47}{20m^3\alpha_{em}}\, .
\eeq
Thus, we have shown that the complete set of the first relativistic
corrections differs essentially from the commonly used term
$\Delta\alpha$. We suppose that for the
electromagnetic polarizabilities of hadrons investigated within the constituent
quark model an analogous  situation may be found.

\section*{Acknowledgments}
A.I.M. thanks the members of the Zweites Physikalisches Institut at
the University of G\"ottingen for the warm hospitality during his stay,
 when a part of this work has been done. One of the authors (M.S.) thanks
Deutsche Forschungsgemeinschaft for the suport of this work
thought he grants Schu222 and 436RUS113/510.

\section*{Appendix A}

In this Appendix we derive the formulas for the result of the action of
the operator $G_0$, \eq{G0}, on the wave function $\ket$, multiplied by some
polynomial of $\vr$. More precisely, we obtain the expression for
$G_0 Y_{lm}(\vr/r) r^n|\psi_0\rangle$ in the form of the product
$Y_{lm}(\vr/r) P(r)|\psi_0\rangle$,where $P(r)$ is some polynomial.
Since the Hamiltonian $H_{nr}$ commutes with the operator of angular momentum
$\vl=\vr\times\vp$, we can make the following transformation:
\beqlab{Y}
G_0 Y_{lm}(\vr/r) r^n\ket=Y_{lm}(\vr/r)
G_0^{(l)}(r^n-\delta_{l\,0}\langle r^n\rangle)\ket\, ,
\eeq
where $G_0^{(l)}=[\varepsilon_0-H_{nr}^{(l)}+i0]^{-1}$, $H_{nr}^{(l)}=-(2\mu
r)^{-1} \partial_r^2 r+l(l+1)/(2\mu r^2)-g/r$ is the radial Hamiltonian with
the angular momentum $l$, and
\beqlab{rn}
\langle r^n\rangle=\langle\psi_0|r^n|\psi_0\rangle=
\frac{(n+2)!}{2^{n+1}}a^{n}\, ,
\eeq
where $a=1/\mu g$.
In the derivation of \eq{Y} we used the identity
\[
(1-\ket\bra)Y_{lm}(\vr/r) r^n \ket= Y_{lm}(\vr/r) (r^n-\delta_{l\,0}\langle
r^n\rangle)\ket\, .
\]
For our purposes it is sufficient to consider the cases $n\geq l-1$
for $l\neq 0$ and $n\geq 1$ for $l=0$. It is easy to check that in
these cases one can represent the result of action of $G_0^{(l)}$ in
r.h.s.  of \eq{Y} in the form
\beq
G_0^{(l)}(r^n-\delta_{l\,0}\langle
r^n\rangle)\ket=\sum_{k=0}^\infty C_k r^k\ket\, ,
\eeq
where $C_k$ are some constants to be found.
Acting on both sides of this equation with the operator
$\varepsilon_0-H_{nr}^{(l)}$ and collecting the coefficients with different
powers of $r$, we obtain
\beqnlab{Glpsi}
r^n-\delta_{l\,0}\langle r^n\rangle&=&-\frac{l(l+1)}{2\mu}C_0
r^{-2}-\frac{(l-1)(l+2)}{2\mu}C_1 r^{-1}+\\
&&+\sum_{k=0}^\infty\left(\frac{(k-l+2)(k+l+3)}{2\mu}C_{k+2}
-g(k+1)C_{k+1}\right)r^k\nonumber\, .
\eeqn
From this relation we can find the coefficients $C_i$. For the
case $n\geq l-1$, $l\neq0$, we finally obtain
\beqlab{G0psi}
G_0 Y_{lm}(\vr/r)
r^n\ket=-Y_{lm}(\vr/r){\frac{\left( n - l + 1 \right) !\, \left( n +
       l + 2 \right) !}{g\,{(2/a)^{n + 1}}\, \left( n + 1 \right)
       !}}\, \sum_{k = l}^{n + 1} \,{\frac{\left( k - 1 \right)
      !\,{(2r/a)^k}} {\left( k - l \right) !\,\left( k + l + 1
         \right) !} }\,\ket \, .  \eeq
For the case $n\geq 1$, $l=0$ we have
\beq
G_0  r^n\ket=-\frac{(n+2)!}{g\,(2/a)^{n+1}}\, \sum_{k = 2}^{n +
1}{\frac1k\left( \frac{(2r/a)^k}{( k + 1 ) !} -{\frac{k + 2}{2}}\right)
}\ket\, .
\eeq
Using the formulas \eq{Glpsi} and \eq{G0psi} one can easily calculate
all matrix elements needed.

\section*{Appendix B}

In this Appendix we check the fulfilment of the low energy
theorem. Namely, we reproduce the two first terms of the expansion with
respect to $\varepsilon_0/M$ of the Compton scattering amplitude at
$\omega=0$:
\beq \label{T00}
T(\omega=0)\approx
-\veps_1\cdot\veps_2^*\frac{(e_1+e_2)^2}{M}
\left(1-\frac{\varepsilon_0}{M}\right).
\eeq
In fact, the first term is contained in \eq{T1}.  In order to
obtain the second term, we have to take into account the
corrections to the current, seagull, wave function, propagator, and
energy due to the Breit Hamiltonian $\tilde{H}_B[\vA]$.
The contribution to the amplitude at $\omega=0$ due to $\vJ_B$ reads
\beqlab{mc0}
T_c=-
\bra [
\veps_2^*\cdot \vJ_B\, G_0\,\veps_1\cdot\vJ_{nr} +
\veps_2^*\cdot \vJ_{nr}\, G_0\,\veps_1\cdot\vJ_B
]\ket +(\veps_1\leftrightarrow \veps_2^*)\, .
\eeq
The contribution due to the correction to seagull (the
terms in $\tilde{H}_B[\vA]$ being  quadratic in $\vA$)  reads
\beqnlab{ms0}
T_s&=&
\bra\left\{
\left(
\frac{e_1^2}{m_1^3}
+\frac{e_2^2}{m_2^3}\right)
\left[
(\veps_1\cdot\vp)(\veps_2^*\cdot\vp)+
(\veps_1\cdot\veps_2^*) \frac{\vp^2}{2}
\right]+ \right.\\
&&
\left.
+\frac{g^2}{m_1m_2}\left[\frac{\veps_1\cdot\veps_2^*}{r}+
\frac{(\veps_1\cdot\vr)(\veps_2^*\cdot\vr)}{r^3}\right]
\right\}\ket \, .\nonumber
\eeqn
The contribution connected with the expansion of propagator with
respect to $H_B$ has the form
\beq \label{mp0}
T_p= -\bra
(\veps_2^*\cdot {\vJ_{nr}}\,G_0 H_B G_0\,\veps_1\cdot {\vJ_{nr}}
\ket +(\veps_1\leftrightarrow \veps_2^*)\, .\nonumber
\eeq
The contribution due to the correction to wave function is
\beqlab{mw0}
T_w= -\bra[
\veps_2^*\cdot {\vJ_{nr}}G_0\veps_1\cdot {\vJ_{nr}}
G_0 H_B+  H_B G_0 \veps_2^*\cdot {\vJ_{nr}}G_0\veps_1\cdot {\vJ_{nr}}
]\ket+(\veps_1\leftrightarrow \veps_2^*) \, .\nonumber
\eeq
At last, the contribution corresponding to the correction to the
ground state energy reads:
\beqlab{me0}
T_e= \delta \varepsilon_0
\langle \psi_0 |
\veps_2^*\cdot {\vJ_{nr}}G_0^2\veps_1\cdot {\vJ_{nr}}
|\psi_0\rangle +(\veps_1\leftrightarrow \veps_2^*) .\nonumber
\end{equation}
Using the results of Appendix A, we obtain the following expressions
for the corrections:
\beqn
T_c&=&-\frac{\veps_1\cdot\veps_2^*}{3}\,g^2\,{\mu }^2\,\left(
\frac{{e_1}}{{m_1}} - \frac{{e_2}}{{m_2}} \right) \, \left[
5\mu\,\left( \frac{{e_1}}{{{m_1}}^3} - \frac{{e_2}}{{{m_2}}^3}
\right)  + 4\frac{{e_1} - {e_2}}{{m_1}\,{m_2}} \right]\, , \\
T_s&=&\frac{\veps_1\cdot\veps_2^*}{6}g^2\,\mu\left[
5\,\mu \, \left( \frac{{{e_1}}^2}{{{m_1}}^3}
+ \frac{{{e_2}}^2}{{{m_2}}^3} \right) +
    \frac{8\,g }{{m_1}\,{m_2}} \right]\, ,\nonumber\\
T_p&=&
\frac{\veps_1\cdot\veps_2^*}{12}
g^2\,{\mu }^3\,{\left( \frac{{e_1}}{{m_1}} - \frac{{e_2}}{{m_2}}
\right) }^2\,
\left[ 7\mu\,\left( \frac1{{m_1}^3} + \frac1{{m_2}^3}
  \right)  + \frac{12}{{m_1}\,{m_2}} \right]\, ,\nonumber\\
T_w&=&
\frac{\veps_1\cdot\veps_2^*}{6}
g^2\,{\mu }^3\,{\left( \frac{{e_1}}{{m_1}} - \frac{{e_2}}{{m_2}}
\right) }^2\,
\left[ 9\mu\,\left( \frac1{{m_1}^3} + \frac1{{m_2}^3}
  \right)  + \frac{14}{{m_1}\,{m_2}} \right]\, ,\nonumber\\
T_e&=&
-\frac{\veps_1\cdot\veps_2^*}{4}
g^2\,{\mu }^3\,{\left( \frac{{e_1}}{{m_1}} - \frac{{e_2}}{{m_2}}
\right) }^2\,
\left[ 5\mu\,\left( \frac1{{m_1}^3} + \frac1{{m_2}^3}
  \right)  + \frac{8}{{m_1}\,{m_2}} \right]\, . \nonumber
\eeqn
Summing up these contributions, we get
\beq
T_B=-\veps_1\cdot\veps_2^*\frac{(e_1+e_2)^2}{2M^2} \mu g^2\, ,
\eeq
which is the second term in \eq{T00}.
Let us consider now the contribution to the Compton amplitude at
$\omega=0$, connected with the spin-dependent terms and the Darwin
terms in Breit Hamiltonian. Note that all these terms are
proportional to either $\delta(\vr)$ or to the operator
$(3n_in_j-\delta_{ij})/r^3$. The terms $\propto\delta(\vr)$ give the
contributions to $T_w$ and $T_e$. Using the results of
Appendix A, it is easy to show that the sum of these two
contributions is zero. The terms $\propto(3n_in_j-\delta_{ij})/r^3$
give the contributions to $T_w$ and $T_p$. Again,
direct calculations show that they also cancel each other. Therefore,
we proved, that the first relativistic correction to the Compton
amplitude at $\omega=0$ is spin-independent, which is in agreement
with the low-energy theorem.

\end{document}